\documentclass{ws-procs975x65}

\begin{document}

\title{Non-Crossing Approximation Study 
of\\MULTI-ORBITAL KONDO EFFECT 
in Quantum Dot Systems
%\footnote{\uppercase{T}his work is supported by etc, etc.}
}

\author{T. KITA, R. SAKANO, T. OHASHI\footnote{\uppercase{P}resent address: \uppercase{C}ondensed-\uppercase{M}atter \uppercase{T}heory \uppercase{L}aboratory, \uppercase{RIKEN}, \uppercase{W}ako, \uppercase{S}aitama 351-0198, \uppercase{J}apan.}, 
and S. SUGA
}

\address{Department of Applied Physics, Osaka University, \\
Suita, Osaka 565-0871, Japan \\
%Condensed-Matter Theory Laboratory, RIKEN, \\
%Wako, Saitama 351-0198, Japan \\
E-mail: kita@tp.ap.eng.osaka-u.ac.jp}

\maketitle

\abstracts{
We study the three-orbital Kondo effect in quantum dot (QD) systems 
by applying the non-crossing approximation 
to the three-orbital Anderson impurity model. 
By investigating the tunneling conductance through a QD, 
we show that the competition between the Hund-coupling and 
the orbital level-splitting gives rise to characteristic behavior 
in transport properties. 
It is found that the Hund-coupling becomes more important 
in the three-orbital case than in the two-orbital case. 
We also show that the enhancement of Kondo temperature 
due to 
the singlet-triplet mechanism suggested for the two-orbital model 
tends to be suppressed 
by the existence of the third orbital. }
%in the three-orbital case. }

\section{Introduction}
Electron transport properties in nanoscale systems 
have been studied extensively. 
In particular, 
recent progress in nanofabrication enables us to 
observe the correlation effect due to the orbital degrees of freedom 
%observe transport properties via new types of Kondo effects 
in highly-symmetric quantum dot (QD) systems. 
In these systems, not only ordinary {\it spin} Kondo effect 
but also various {\it orbital} Kondo effects, 
such as SU(4) Kondo effect in a carbon nanotube QD,\cite{jari05} 
singlet-triplet Kondo effect,\cite{sas00} etc.,
have been observed. 
These observations of orbital Kondo effects in QD systems 
have activated theoretical works for 
%transport properties via 
orbital Kondo effects.  
Theoretical studies using the two-orbital Anderson impurity model (AIM)
\cite{eto05,pust01,hofs02,izum01,choi05}
have pointed out the importance of orbital degeneracy;  
(i) When one electron occupies two nearly-degenerate orbital-levels, 
the Kondo temperature $T_K$ gets enhanced 
as the system approaches to the SU(4) symmetric point 
where two orbitals are degenerate. 
(ii) When two electrons occupy two orbital-levels, 
the ground state of the isolated Anderson impurity 
from conduction electrons is the triplet (spin $S=1$) 
for degenerate energy levels of two orbitals. 
When the level-splitting $\Delta\varepsilon$ becomes larger, 
the ground state changes into the singlet. 
As $\Delta\varepsilon$ increases, 
$T_K$ takes a maximum at the point where 
the energy levels of the two states are degenerate. 

In contrast to the detailed investigation of the two-orbital case, 
the three-orbital Kondo effect in QD systems 
has not been sufficiently understood yet. 
Experimentally, the three-orbital Kondo effect has been realized 
in the vertical QD,\cite{taru00}
where the orbital degeneracy is well controlled by 
an external magnetic field or deformation of the QD.\cite{toku01} 
Therefore, it is desirable to study transport properties via 
the three-orbital Kondo effect by systematically changing orbital degeneracy
in the three-orbital AIM. 

In this paper, we study the three-orbital Kondo effect 
by applying the non-crossing approximation (NCA)\cite{bick87} 
to the three-orbital AIM. 
We focus on the Kondo effect for integer filling; 
two or three electrons occupy three orbitals. 
By investigating the tunneling conductance through a QD, 
we show that the competition between the Hund-coupling and 
the orbital level-splitting gives rise to characteristic transport properties. 
It is found that the Hund-coupling becomes more important 
in the three-orbital case than in the two-orbital case. 
We also show that the enhancement of $T_K$ due to 
the singlet-triplet mechanism suggested for the two-orbital model 
tends to be suppressed 
%in the three-orbital case. 
by the existence of the third orbital.

This paper is organized as follows. 
In the next section, we briefly mention the model and method. 
In Sec. 3, we show the numerical results, and 
discuss the characteristic transport properties 
due to the three-orbital Kondo effect 
in comparison with the two-orbital case. 
Brief summary is given in Sec. 4.

\section{Model and Method}
We study the three-orbital Kondo effect 
by exploiting the three-orbital AIM, 
\begin{eqnarray}
H &=& H_{\textrm{c}}+H_{\textrm{loc}}+H_{\textrm{mix}}, \label{eq:model} \\
H_{\textrm{c}} &=& \sum _{ k i \sigma } \varepsilon_{k}c_{k i \sigma }^{\dagger }c_{k i \sigma}, \nonumber \\
H_{\textrm{loc}} &=& \sum _{i \sigma}E_{d_i} d_{i \sigma}^{\dagger }d_{i \sigma}+U \sum _{i}n_{d_{i \uparrow }}n_{d_{i \downarrow }}
+U^{\prime} \sum _{(i\neq j) \sigma \sigma^{\prime} }n_{d_{i \sigma}}n_{d_{j \sigma^{\prime}  }}
-J\sum _{(i\neq j)} \textbf{S}_{d_i} \cdot \textbf{S}_{d_j}, \nonumber \\
H_{\textrm{mix}} &=& \sum _{ k i } V_{ki} 
                     \left(c_{k i \sigma }^{\dagger }d_{i \sigma}+H.c. \right),
                     \nonumber
%\label{eq:model}
\end{eqnarray}
where $H_{\textrm{c}}$, $H_{\textrm{loc}}$, and $H_{\textrm{mix}}$ describe 
a part of conduction electrons in the leads, a QD, 
and mixing between the leads and the QD, respectively.  
Here, $c_{k i \sigma } (d_{i \sigma})$ annihilates a conduction electron 
(localized electron in the QD) with spin $\sigma$ in the orbital $i$, 
and $n_{d_{i\sigma}}=d_{i \sigma}^\dag d_{i \sigma}$. 
$\textbf{S}_{d_i}$ is the spin operator for a localized electron in the orbital $i$.
$E_{d_i}$ denotes the local level of the orbital $i$, 
$U (U^\prime)$ the intraorbital (interorbital) Coulomb interaction 
and $J$ denotes the Hund-coupling among orbitals. 

To analyze our model (\ref{eq:model}), we use the NCA.\cite{bick87} 
By calculating the local density of states 
%for spin $\sigma$ and orbital $i$ 
$\rho _{{i \sigma}}(\omega)$, 
we obtain the tunneling conductance $G$ through the QD,\cite{meir92} 
\begin{equation}
G=\frac{e^2 \Gamma}{\hbar } 
\int d \omega \left(-\frac{df(\omega)}{d \omega} \right) 
\sum _{i \sigma} \rho _{i \sigma}(\omega), 
\label{G}
\end{equation}
where $\Gamma$ denotes the strength of the hybridization 
between conduction electrons and localized electrons in the QD, 
and $f(\omega)$ is the Fermi distribution function. 

\section{Numerical Results}
%In this section, we discuss our numerical results for the tunneling conductance. 
%In the next subsection \ref{subsec:three}, 
We address the case where three electrons occupy three orbitals in the QD 
(referred as the case of three electrons in three orbitals), 
and the case where two electrons occupy three orbitals in the QD 
(referred as the case of two electrons in three orbitals). 
%in the subsection \ref{subsec:two}. 
We especially focus on the competition between the 
Hund-coupling $J$ and the orbital level-splitting $\Delta\varepsilon$. 
We set $U=U^\prime$ in the following calculation and 
use $\Gamma$ in units of the energy. 
\begin{figure}[bth] 
\begin{center}
\epsfxsize=9cm   %width of figure - will enlarge/reduce the figures
\epsfbox{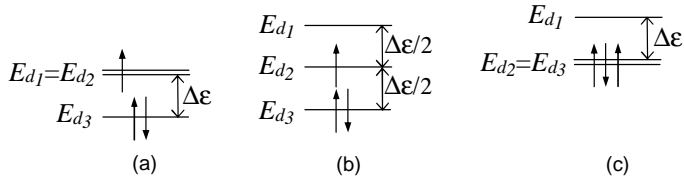}
\caption{Three types of the orbital splitting. 
(a) $E_{d_1}=E_{d_2}=E_{d_3}+\Delta\varepsilon$, 
(b) $E_{d_1}=E_{d_2}+\Delta\varepsilon/2, E_{d_3}=E_{d_2}-\Delta\varepsilon/2$, 
(c) $E_{d_1}=E_{d_2}+\Delta\varepsilon=E_{d_3}+\Delta\varepsilon$. 
%$\Delta\varepsilon$ is set without changing the electron filling in the QD. 
The electron filling is assumed to be unchanged due to $\Delta\varepsilon$. 
\label{3o3e}}
\end{center}
\end{figure}

\begin{figure}[tbh]
\begin{center}
\epsfxsize=12.5cm   %width of figure - will enlarge/reduce the figures
\epsfbox{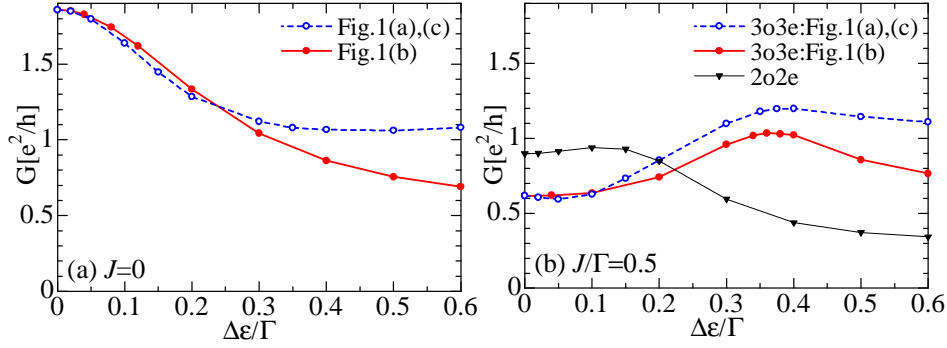}
\caption{The conductance $G$ as a function of the level-splitting $\Delta \varepsilon$ 
for the case of three electrons in three orbitals 
(a) without Hund-coupling $J=0$, 
(b) with Hund-coupling $J/\Gamma=0.5$. 
Parameters are set as $U/\Gamma=U^\prime/\Gamma=10$ and $T/\Gamma=0.05$. 
The types of the orbital splitting as shown in 
Fig.~\ref{3o3e} (a) and (c) (dashed line), and 
Fig.~\ref{3o3e} (b) (thick solid line) are considered. 
For comparison, we plot the result for the case of two electrons in two orbitals 
for the same parameters (thin solid line). 
 \label{3o3eJ0}}
\end{center}
\end{figure}

%\subsection{Three Electrons in Three Orbitals \label{subsec:three}}
We first investigate the orbital Kondo effect for 
the case of three electrons in three orbitals. 
We consider three types of the orbital splitting 
shown in Fig.~\ref{3o3e}. 
In large $\Delta\varepsilon$ limit 
for each type of the orbital splitting, it is expected that 
(a) SU(4) Kondo effect, (b) SU(2) Kondo effect, 
and (c) SU(4) Kondo effect with three electrons 
are realized, respectively. 
%\paragraph{The results without Hund-coupling ($J=0$).}
Let us start our discussion for the results without Hund-coupling. 
In Fig.~\ref{3o3eJ0}(a), we plot the tunneling conductance $G$ 
as a function of $\Delta\varepsilon$ 
for each type of the orbital splitting in Fig.~\ref{3o3e}. 
The conductance of (a) and (c) types results in the same behavior 
because of particle-hole symmetry. 
For each type of the orbital splitting, 
$G$ monotonically decreases as $\Delta\varepsilon$ increases. 
The reduction of the orbital degeneracy due to $\Delta\varepsilon$ 
lowers $T_K$, 
which yields the decrease in $G$. 
%which decreases the tunneling conductance 
%via the Kondo resonance at finite temperature. 
For large $\Delta\varepsilon$ ($\Delta\varepsilon/\Gamma > 0.4$), 
$G$ of (a) and (c) types is larger than $G$ of (b) type. 
This behavior is consistent with the fact that 
$T_K$ in SU(4) Kondo effect is higher than $T_K$ in SU(2) Kondo effect, 
although $G$ in both cases take the same value at zero temperature.\cite{saka06} 

%

%\paragraph{The results including the effect of Hund-coupling ($J\neq0$).}
We next show the results including the effects of the Hund-coupling $J$ in Fig.~\ref{3o3eJ0} (b). 
For comparison, we plot the result for the case of two electrons in two orbitals. 
Each conductance exhibits a maximum structure, which is due to 
the competition between the Hund-coupling and the level-splitting. 
\begin{figure}[b]
\begin{center}
\epsfxsize=12.5cm   %width of figure - will enlarge/reduce the figures
\epsfbox{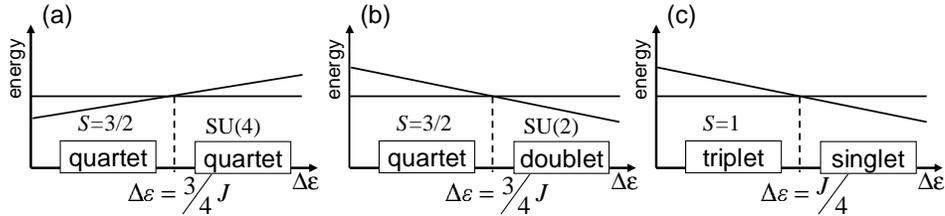}
\caption{Schematic diagram for the change of the ground state 
%This is the consideration under non-interacting 
of $H_{\textrm{loc}}$ 
with finite $J$ and $\Delta\varepsilon$. 
(a) Orbital splitting is introduced as shown in Fig.~\ref{3o3e} (a). 
(b) Orbital splitting is introduced as shown in Fig.~\ref{3o3e} (b). 
%(a) with orbital splitting shown in Figure~\ref{3o3e} (a). 
%(b) with orbital splitting shown in Figure~\ref{3o3e} (b). 
(c) The case of two electrons in two orbitals. 
\label{GS}}
\end{center}
\end{figure}
For detailed explanation, we describe the ground state of $H_{\textrm{loc}}$. 
For the finite Hund-coupling $J$, 
the ground state is the spin $S=3/2$ state (quartet state) 
at $\Delta\varepsilon=0$. 
When the orbital splitting is introduced 
as shown in Fig.~\ref{3o3e}(a), 
the ground state changes into the quartet state 
with fourfold degeneracy of the spin and orbital degrees of freedom 
at large $\Delta\varepsilon$, 
where the SU(4) Kondo effect is induced, as mentioned above. 
On the other hand, when the orbital splitting is introduced 
as shown in Fig.~\ref{3o3e}(b), 
the ground state changes from the $S=3/2$ quartet state to 
the doublet state with the spin $\uparrow$ and $\downarrow$ degrees of freedom, 
which leads to the ordinary SU(2) Kondo effect. 
At the critical point where the ground state changes 
as shown in Fig.~\ref{GS}, 
the degeneracy of the ground state is enlarged, 
which gives rise to the enhancement of the Kondo temperature. 
Actually, as $\Delta\varepsilon$ increases, 
the conductance $G$ in Fig.~\ref{3o3eJ0}(b) take a maximum 
near the critical point. 
Note that $G$ for the case of Fig.~\ref{3o3e}(a) is somewhat larger than 
that of Fig.~\ref{3o3e}(b) near the critical point, because 
the degeneracy for the case of Fig.~\ref{3o3e}(a) is larger than 
that of Fig.~\ref{3o3e}(b). 
The mechanism of the maximum structure of $G$ 
is similar to the singlet-triplet Kondo effect, 
which occurs in the case of two electrons in two orbitals. 
The difference appears at the point where $T_K$ is enhanced; 
In the three-orbital case, $T_K$ is enhanced at 
$\Delta\varepsilon\sim\frac{3}{4}J$, while 
$T_K$ is enhanced at $\Delta\varepsilon\sim\frac{1}{4}J$ 
in the two-orbital case. 
This indicates that the effect of $J$ remains 
even at larger $\Delta\varepsilon$ in the three-orbital case, 
because three electrons gain more Hund-coupling energy than two electrons. 
Namely, the Hund-coupling becomes more important 
in the three-orbital case than in the two-orbital case. 

\begin{figure}[ht]
\begin{center}
\epsfxsize=5cm   %width of figure - will enlarge/reduce the figures
\epsfbox{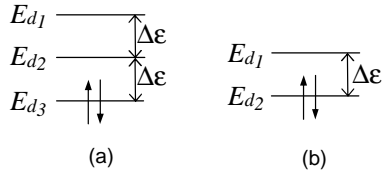}
\caption{(a) Level-splitting of three orbitals. 
$E_{d_1}=E_{d_2}+\Delta\varepsilon, 
 E_{d_3}=E_{d_2}-\Delta\varepsilon$. 
(b) Level-splitting of two orbitals. 
$E_{d_1}=E_{d_2}+\Delta\varepsilon$. 
The electron filling is assumed to be unchanged due to $\Delta\varepsilon$. 
%$\Delta\varepsilon$ is set without changing the electron filling in the QD. 
\label{3o2e}}
\end{center}
\end{figure}

%\subsection{Two Electrons in Three Orbitals \label{subsec:two}}
We now turn to the case of two electrons in three orbitals. 
Here, we discuss how the third orbital affects the transport properties, 
by comparing the results to those for the case of two electrons in two orbitals. 
We consider the level-splitting $\Delta\varepsilon$ as shown in Fig.~\ref{3o2e}.
\begin{figure}[b] 
\begin{center}
\epsfxsize=12.5cm   %width of figure - will enlarge/reduce the figures
\epsfbox{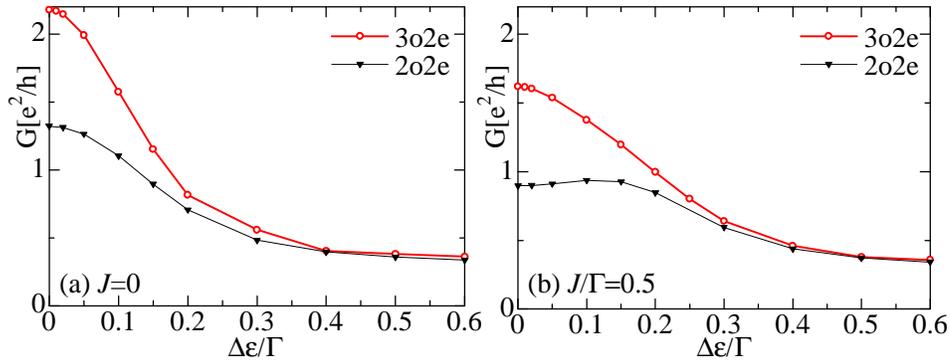}
\caption{The conductance $G$ for the case of two electrons in three orbitals (3o2e: thick solid line) 
and the case of two electrons in two orbitals (2o2e: thin solid line) 
as a function of level-splitting $\Delta\varepsilon$ 
for $U/\Gamma=U^\prime/\Gamma=10,T/\Gamma=0.05$. 
We show the results (a) without Hund-coupling ($J=0$) and  
(b) with finite Hund-coupling $J/\Gamma =0.5$. 
 \label{3o2eJ0fJ}}
\end{center}
\end{figure}
%
%\paragraph{The results without Hund-coupling ($J=0$).}
In Fig.~\ref{3o2eJ0fJ}(a), 
we plot the tunneling conductance $G$ as a function of 
the level-splitting $\Delta\varepsilon$ for 
the case of two electrons in three orbitals and 
the case of two electrons in two orbitals without Hund-coupling. 
As $\Delta\varepsilon$ increases, 
$G$ in both cases decreases monotonically, 
and approaches the same value. 
At large $\Delta\varepsilon$ ($\Delta\varepsilon/\Gamma>0.4$), 
the orbital with the highest energy-level does not contribute to the conductance 
so that $G$ results in the same value. 

%\paragraph{The results including the effect of Hund-coupling ($J\neq0$).}
For the finite Hund-coupling $J/\Gamma=0.5$, 
we find the noticeable difference 
between the two- and three-orbital cases. 
As shown in Fig.~\ref{3o2eJ0fJ}(b), 
for the two-orbital case, 
the Hund-coupling $J$ leads to a hump structure 
with a maximum around $\Delta\varepsilon/\Gamma \sim 0.12$, 
which is due to the singlet-triplet Kondo effect. 
On the other hand, for the three-orbital case, 
the conductance $G$ is not enhanced but monotonically decreases, 
as $\Delta\varepsilon$ increases. 
For small $\Delta\varepsilon$, the ground state of $H_{loc}$ is the triplet state, 
and then it changes into the singlet state. 
Therefore, it is expected that $G$ gets enhanced due to the singlet-triplet Kondo effect 
at intermediate $\Delta\varepsilon$. 
However, for the three-orbital case without $\Delta\varepsilon$, 
the ground state of $H_{loc}$ has ninefold degeneracy, 
which gives very high Kondo temperature. 
In this case, for small $\Delta\varepsilon$, 
the naively expected $S=1$ Kondo effect is not realized 
but the Kondo effect with high $T_K$ similar to 
that without $\Delta\varepsilon$ occurs. 
Therefore, the enhancement of $T_K$ due to the singlet-triplet mechanism 
merges into decrease of $T_K$ due to the collapse of 
the ninefold degenerate Kondo effect, 
which makes difficult to see the singlet-triplet Kondo effect. 

\section{Summary}
We have studied the three-orbital Kondo effect 
in QD systems by exploiting 
the three-orbital AIM. 
By means of NCA, 
we have calculated the tunneling conductance through the QD. 
We have found that the Hund-coupling becomes more important 
in the three-orbital case than in the two-orbital case. 
We have also shown that the enhancement of $T_K$ due to 
the singlet-triplet mechanism tends 
to be suppressed by the existence of the third orbital. 
%These behaviors of the tunneling conductance may be observed experimentally 
%in the near future. 
It is expected that 
the characteristic behavior of the tunneling conductance obtained here 
will be observed experimentally in the near future. 

\section*{Acknowledgments}
The authors thank S. Amaha for valuable discussions. 
This work was partly supported by a Grant-in-Aid from the Ministry 
of Education, Science, Sports and Culture of Japan.

\end{document}